# Further Analysis on the Mystery of the Surveyor III Dust Deposits

John Lane[3], Steven Trigwell[2], Paul Hintze[1], Philip Metzger[1]

[1] Granular Mechanics and Regolith Operations Lab, NASA, Kennedy Space Center, FL 32899
[2] Applied Technology, Siera Lobo-ESC, Kennedy Space Center, FL 32899
[3] Granular Mechanics and Regolith Operations, Easi-ESC, Kennedy Space Center, FL  32899

## ABSTRACT

The Apollo 12 lunar module (LM) landing near the Surveyor III spacecraft at the end of 1969 has remained the primary experimental verification of the predicted physics of plume ejecta effects from a rocket engine interacting with the surface of the moon.  This was made possible by the return of the Surveyor III camera housing by the Apollo 12 astronauts, allowing detailed analysis of the composition of dust deposited by the LM plume.  It was soon realized after the initial analysis of the camera housing that the LM plume tended to remove more dust than it had deposited.  In the present study, coupons from the camera housing have been reexamined.  In addition, plume effects recorded in landing videos from each Apollo mission have been studied for possible clues. Several likely scenarios are proposed to explain the Surveyor III dust observations.  These include electrostatic levitation of the dust from the surface of the Moon as a result of periodic passing of the day-night terminator; dust blown by the Apollo 12 LM flyby while on its descent trajectory; dust ejected from the lunar surface due to gas forced into the soil by the Surveyor III rocket nozzle, based on Darcy's law; and mechanical movement of dust during the Surveyor landing.  Even though an absolute answer may not be possible based on available data and theory, various computational models are employed to estimate the feasibility of each of these proposed mechanisms.  Scenarios are then discussed which combine multiple mechanisms to produce results consistent with observations.

## LUNAR SURVEYOR III SAMPLES

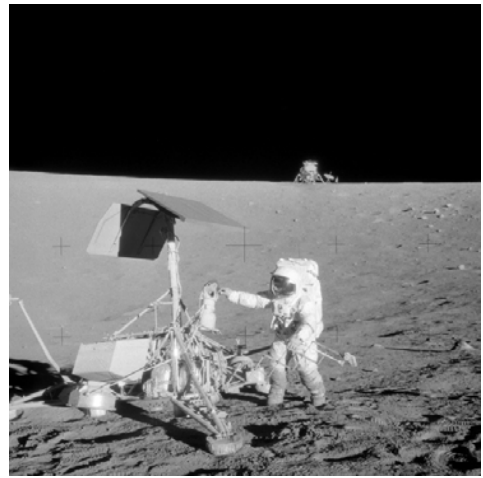

Figure 1: Alan Bean examining Surveyor III.  Note that the Apollo 12 LM is in the background.

The Surveyor III parts brought back to Earth by Apollo 12 (see Figure 1) are currently stored at the Johnson Space Center (JSC) Lunar Sample Curation and information on the chemical analyses is found at the Lunar and Planetary Institute.  Several of the parts were obtained for analysis in the present study (see and Figures 2, 3, and 5).  Those parts are summarized in Table 1.

## LOFTING BY APOLLO 12 FLYBY

First we investigate whether the flyby of the Apollo 12 LM during descent could have sprayed dust onto the Surveyor III spacecraft.  The flyby occurred at an altitude of approximately 67 m (see Figure 4).

**Shear Stress Simulations of LM Engine Plume Flyby**

Starting with a Fluent CFD simulation of the Apollo LM engine in a lunar-like environment (background pressure is artificially set to a small non-zero value in order to achieve convergence), three gas parameters are computed for every point in a 2D non-uniform spatial grid: gas density $\rho(k)$, gas temperature $T(k)$, and gas velocity $\mathbf{v}(k)$.  Each CFD simulation and computed gas parameter output set corresponds to a specific engine height $h$ above the lunar surface.   The CFD simulation generates gas parameter data at specific spatial points corresponding to the $x$-$y$ coordinates contained in the grid point array, $\mathbf{r}(k)$.  Note that for the Fluent CFD cases considered in this report, vertical positions are described by the coordinate $x$ and horizontal positions are described by the coordinate *y*. Since the CFD generation of spatial points is based on algorithms which are used to minimize error in partial differential equations describing the laws of fluid mechanics, the grid points for all practical purposes are randomly distributed. Therefore, finding a specific point nearest a field point *x-y* and its nearest neighbors, involves searching the entire **r**(*k*) array for $k = 1... N$.

**Table 1. Surveyor III Parts Under Investigation.**

| Type of Sample | Part Designation | Comments |
|---|---|---|
| Cutouts from the Surveyor III Camera | JPL #1037 | Sample from the camera bottom. |
| | JPL #2048 | Flat side of camera. |
| | JPL #2049 | Sample from camera pointing away from the Apollo 12 LM. |
| | JPL #2050 (JSC #3160026) | Sample from camera pointing towards the Apollo 12 LM. |
| | JPL #2051 (JSC #3160027) | Sample from camera pointing towards the Apollo 12 LM. |
| | JPL #2052 | Sample from camera pointing away from the Apollo 12 LM. |
| Acetate Tape Samples of the Regolith Material (collected from underneath the camera clamp) | 327 082    O52 | |
| | 327 083    O53 | |
| Section of Cylindrical Rod (strut for the radar) | Sample from each end | |

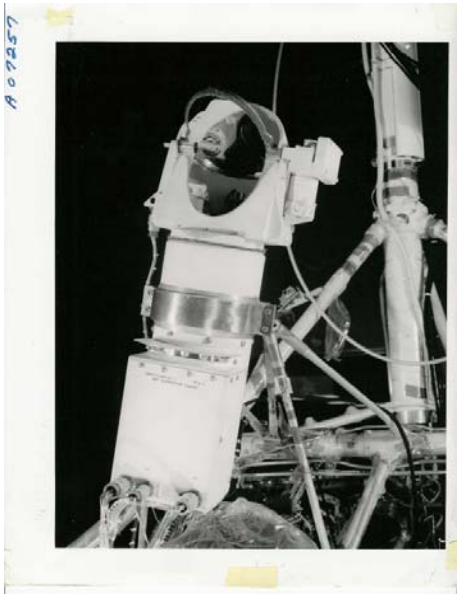

Figure 2: Surveyor III camera on the moon.

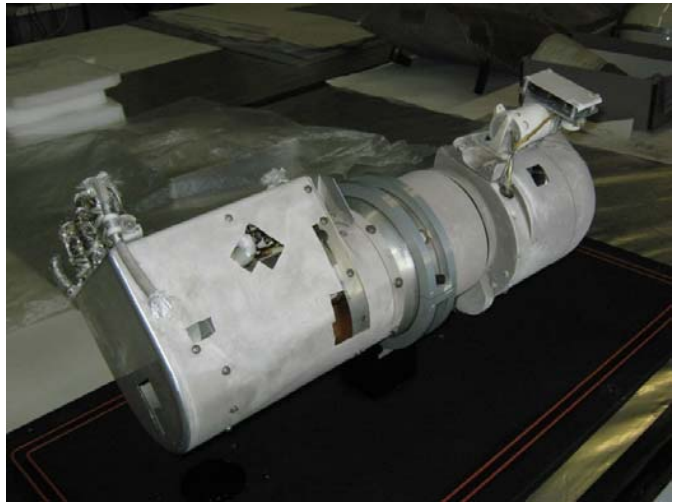

**Figure 3: The camera module showing the bottom cutout**.

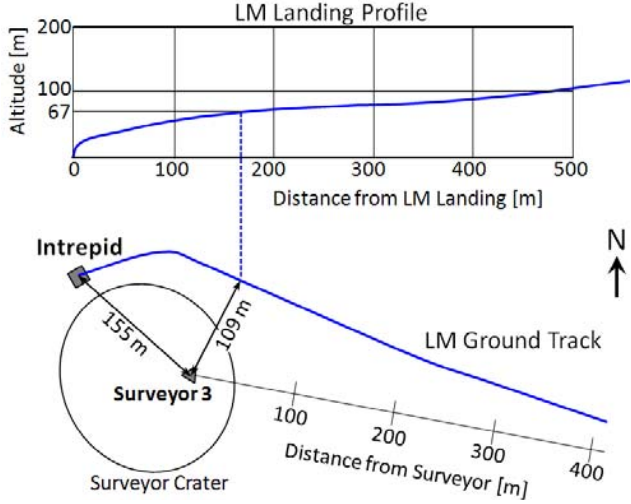
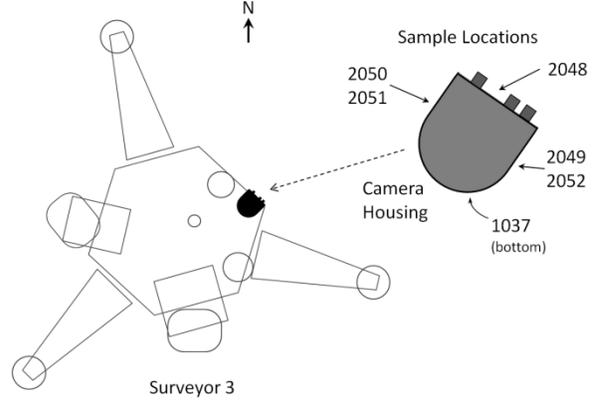

Figure 4: Profile and ground track of Apollo 12 LM approach to landing site.

Figure 5: Sample locations on camera housing.

To compute the shear stress from the CFD output, the grid data is resampled along the lunar surface boundary at $\mathbf{r}_{mn} = (m\Delta x, n\Delta y)$ for $m = 0, 1, 2$ and $n = 0 \ldots N_x$. The shear stress is defined by:

$$\tau \equiv \mu \frac{\partial v_y}{\partial x} , \qquad (1)$$

where $\mu$ is the dynamic viscosity of the gas, $v_y$ is the horizontal component of the gas velocity along the horizontal surface boundary, and $x$ is the distance above the surface. The gas dynamic viscosity is a function of temperature and can be computed by *Sutherland's formula* approximation:

$$\mu(T) \equiv \mu_0 \frac{T_0 + C}{T + C} \left(\frac{T}{T_0}\right)^{3/2} , \qquad (2)$$

where the parameters $\mu_0$, $C$, and $T_0$ are dependent on the gas composition. A discrete approximation to the shear stress of Equation (1) can be computed using the resampled CFD output:

$$\tau_n \equiv \mu_n \frac{v_{y\,2,n} - v_{y\,0,n}}{2\Delta x} , \qquad (3)$$

where

$$\mu_n \equiv \mu_0 \frac{T_0 + C}{T_{1,n} + C} \left(\frac{T_{1,n}}{T_0}\right)^{3/2} . \qquad (4)$$

An equivalent shear velocity (sometimes referred to as *saltation velocity*) can be computed from Equation (3) and the resampled gas density:

$$u_n \equiv \left(\frac{\tau_n}{\rho_{1,n}}\right)^{1/2} . \qquad (5)$$

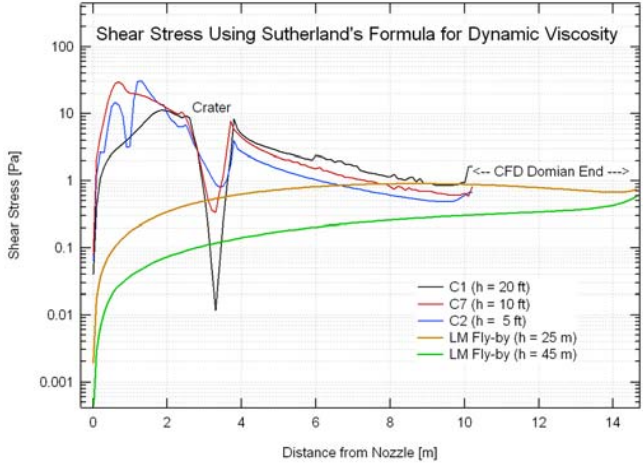

**Figure 6: Shear stress along the surface for five cases generated by Fluent CFD: *h* = 5, 10, 20, 25 [ft] and 45 [m].**

Shear stress along the surface has been computed for the five cases generated by Fluent CFD: *h* = 5, 10, and 20 [ft] and *h* = 25 and 45[m]. The shear stress is computed and plotted for these five cases in Figure 6. The shear stress in this figure was generated with $\Delta x = 0.001$ [m] in Equation (3).

**Threshold Shear Stress Velocity**

Starting with the approach of Sagan (1990), the threshold shear stress velocity is:

$$u^* = \frac{1}{D}\left|\alpha_0 \eta - \alpha_1 \frac{\eta^2}{\Gamma} - \alpha_4 \Gamma\right| \qquad , \qquad (6)$$

where,

$$\Gamma \equiv \left(\phi_2 + (\phi_2^2 - \phi_1^2)^{1/2}\right)^{1/3} \qquad , \qquad (7)$$

and,

$$\phi_1 \equiv \alpha_2 \eta^3 \qquad \phi_2 \equiv \alpha_3 \beta D^{1/2} \eta \qquad . \qquad (8)$$

The gas kinematic viscosity is the gas dynamic viscosity divided by its density: $\eta = \mu/\rho$. With Equation (6) and its variables expressed in SI units, the constant coefficients are: $\alpha_0 = 2/15$; $\alpha_1 = 8.3995$; $\alpha_2 = 250000$; $\alpha_3 = 843750$; $\alpha_4 = 0.0021165$.

The result in Sagan (1990) was greatly simplified with the assumption that the Reynolds number is much less than 1: $Re = u^* D/\eta \ll 1$, or $u^* \ll \eta/D$. Equations (6) through (8) do not make that assumption. The threshold shear stress velocity of Equation (6) can be converted into a threshold shear stress, similar to that of Equation (5):

$$\tau^* \equiv \rho (u^*)^2 \qquad . \qquad (9)$$

The cohesion force (*interparticle force*) in Sagan (1990) is approximated as:

$$I_p \equiv \beta D^{1/2} \qquad , \qquad (10)$$

where again, Equation (10) and its variables are all expressed in SI units. Sagan (1990) used a value of $\beta = 6 \times 10^{-7}$ to represent the cohesive force of particles on the surface of Triton. To simulate a zero cohesion force, $\beta$ is set equal to zero.

Figure 7 shows the trajectory simulation results: the radial distance traveled by particle from ground track position as a function of particle diameter. Circles represent the distances traveled by particles that began their trajectories at different starting points, both *x* and *y* (relative to the impingement point on the centerline of the LM engine). Radial distance from the centerline is *x*.

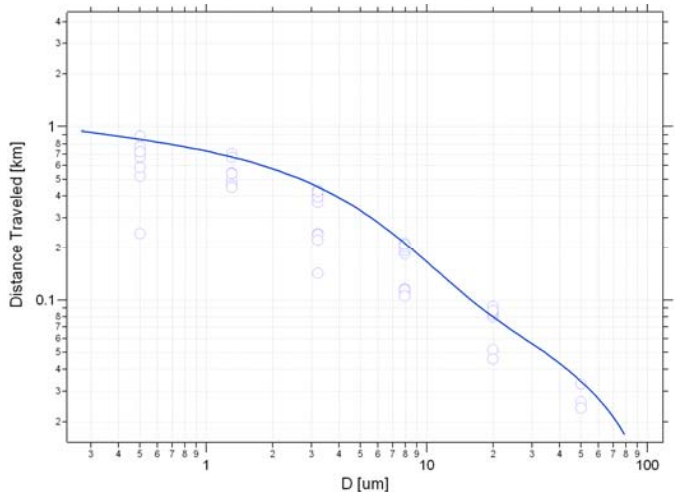

**Figure 7: Radial distance traveled by particle from ground track position as a function of particle diameter for $h = 45$ m. Circles represent different starting points, both $x$ and $y$. Solid line is the average maximum value of the individual trajectories.**

Height above the surface is $y$. Note that particle trajectories start at non-zero values in $y$ (slightly above the lunar surface) to represent the effects of the collisional regime of particles rolling on the surface, which bounces them slightly above the ground until the point where aerodynamic forces dominate. Explicitly modeling the collisional regime flow on the surface for a realistic assemblage of lunar soil particles is beyond the current state of the art, but simply beginning particles a short distance above the ground as employed here has been validated through comparison with photogrammetry of the Apollo landing videos to produce good approximation in the simulated trajectories (Lane et al, 2008; Lane et al, 2010; Lane and Metzger, 2011; Immer et al, 2011). The solid line in Figure 7 is a smoothed curve of the maximum values of the individual trajectories. Note that there is an abrupt cut-off as the particle size approaches 100 µm.

Figure 8 shows post-processed results of a Fluent CFD simulation corresponding to the Apollo 12 LM flyby at $h = 45$ [m]. Note that the actual height of the closest approach distance from the LM ground track to the Surveyor III site is approximately $h = 65$ [m]. ($h = 65$ [m] Fluent CFD results could not be obtained during the project period, possibly due to numerical convergence problems resulting from the larger solution domain with low background pressure representing lunar vacuum.) Referring to Figure 8, the left axis and blue line represents the radial distance traveled by a particle from the ground track position. The shaded portion represents all particles whose horizontal trajectory distance is equal to or greater than the distance to the Surveyor spacecraft ($R = 109$ [m]). The right axis corresponds to the difference in shear stress, Equation (3) and threshold shear stress, Equation (9), indicating the region where lift may occur without the need of particle collisions to initiate a lift process. The zero cohesion force case is shown by the green line and green shading. Sagan's cohesion force, using $\beta = 6\times10^{-7}$, is shown by the brown line and shading.

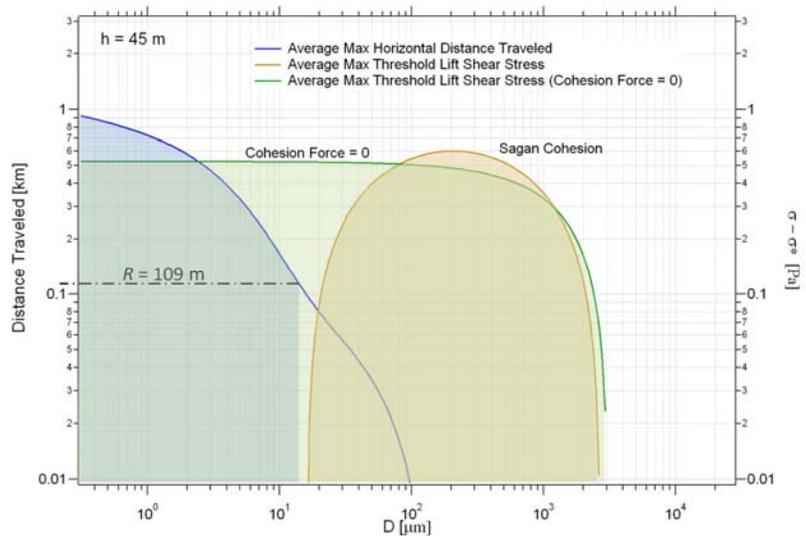

**Figure 8: LM Flyby simulation at $h = 45$ [m]. Left axis: (blue line) radial distance traveled by particle from ground track position. Right axis: (green and brown lines) region where shear stress is greater than threshold shear stress, resulting in particle lift for zero cohesion force (green line) and for Sagan's cohesion force (brown line).**

Based on particle trajectory simulations for the $h = 45$ m case and assuming the zero cohesion case, dust reaches the Surveyor III site from the LM flyby closest approach ($R = 109$ m). Particle sizes up to $D = 13$ um reach the Surveyor III with velocities up to 130 m/s. Particles sizes $D > 13$ um, are also ejected in the zero cohesion case, but fall back to the surface before traversing the complete 109 m distance to the Surveyor. However, it is known that cohesion in the lunar soil is significant (Walton, 2008). Using Sagan's cohesion force, particles in the size range of $17 < D < 2600$ µm can be lifted by the gas shear stress, based on the $h = 45$ m Fluent CFD case. According to the plots in Figure 8, there is no particle size range that shows an overlap between the trajectory distance and threshold shear stress for $R = 109$ m. Inspecting the plot in more detail, particles of $D = 20$ µm would reach a distance of $R = 80$ m. Thus, all the particles lifted by the shear stress against Sagan's cohesion shear stress fail to reach the Surveyor III spacecraft.

Sagan's cohesion force (Iversen,1982) predicts interparticle cohesion forces on the order of nN for particles in the range of 10 – 100 µm. Sagan's cohesion force is about a 1000 times smaller than the cohesion force predicted by Walton (2008), which predicts µN particle pull-off forces. Even for the small Sagan cohesion force (~ nN for 10-100 µm particles), particles do not make it to the Surveyor III site, unless the cohesion force is much smaller (maybe a factor of 10 smaller would do it). But a factor of 1000 larger would certainly decrease the chance of particle spray from the LM flyby.

**DARCY'S LAW**

Darcy's Law describes the volume flow rate $Q$ of a gas or liquid of viscosity $\mu$ through a solid porous medium of permeability $\kappa$, due to a pressure gradient over a length $\Delta L$:

$$Q = -\frac{\kappa A}{\mu}\frac{\Delta P}{\Delta L} \quad , \tag{11}$$

where $A$ is the cross-sectional area of the flow volume and $\Delta P/\Delta L$ is the pressure gradient. Darcy's Law can be used to describe the portion of gas that is injected into the soil immediately beneath the rocket nozzle. Since $Q$ in Equation (11) can be replaced by the gas velocity $v(t)$ scaled by $A$, the initial velocity at $t = 0$ is:

$$v_0 = v(0) = -\frac{\kappa}{\mu}\frac{\Delta P_0}{\Delta L} \quad , \tag{12}$$

Immediately before engine cutoff, the pressure on the surface due to the lander's rocket gas impingement at $t = 0$ can be approximated as:

$$\Delta P_0 = \frac{Mg}{A} \quad , \tag{13}$$

where $M$ is the mass of the lander, $g$ is acceleration due to lunar gravity, and $A$ is the area on the surface over which the pressure is acting. Combining Equations (12) and (13):

$$v_0 = -\frac{\kappa M g}{\mu \Delta L A} \quad , \tag{14}$$

The trajectory of regolith particles lifted upwards from the surface after engine cutoff due to the release of plume gas trapped in the soil can be computed by considering Newton's second law of motion, $F = ma$. $F$ for can be expressed as the gas pressure at the surface due to the trapped gas below as, $A\rho v^2(t)/2$, where $\rho$ is the gas density. The term $ma$ can be expressed as $(\rho A L) dv(t)/dt$, which then leads to:

$$\frac{dv(t)}{dt} = -\frac{v(t)^2}{2L} \qquad (15)$$

where the minus sign is needed because of velocity decreasing. The solution to Equation (15) is:

$$v(t) = \frac{v_0}{v_0 t/2L + 1} \qquad (16)$$

Equation (15) is an approximation the velocity of a particle lifted from the surface due to the Darcy effect, immediately following engine cutoff.

A single particle trajectory can be obtained by computing the position as a function of time by evaluating the integral of velocity from 0 to $t$:

$$y(D,t) = \int_0^t \big(v(t') - v_T(D)\big) dt' \qquad (17)$$

where $y(D, t)$ is the vertical position of the particle and $v_T(D)$ is its terminal velocity in the gas flow. Note the particle diameter dependence of the terminal velocity term. The above integral can be evaluated as:

$$y(D,t) = 2L \ln(v_0 t/2L + 1) - v_T(D) t \qquad (18)$$

The terminal velocity term is due to the drag on the particle by the gas and the pull of lunar gravity. Since the gas density is very small as compared to the particle density $\rho_p$, the terminal velocity $v_T(D)$ depends only on gas viscosity $\mu(T)$, as given by Equation (2), as well as gravity $g$ and particle diameter $D$:

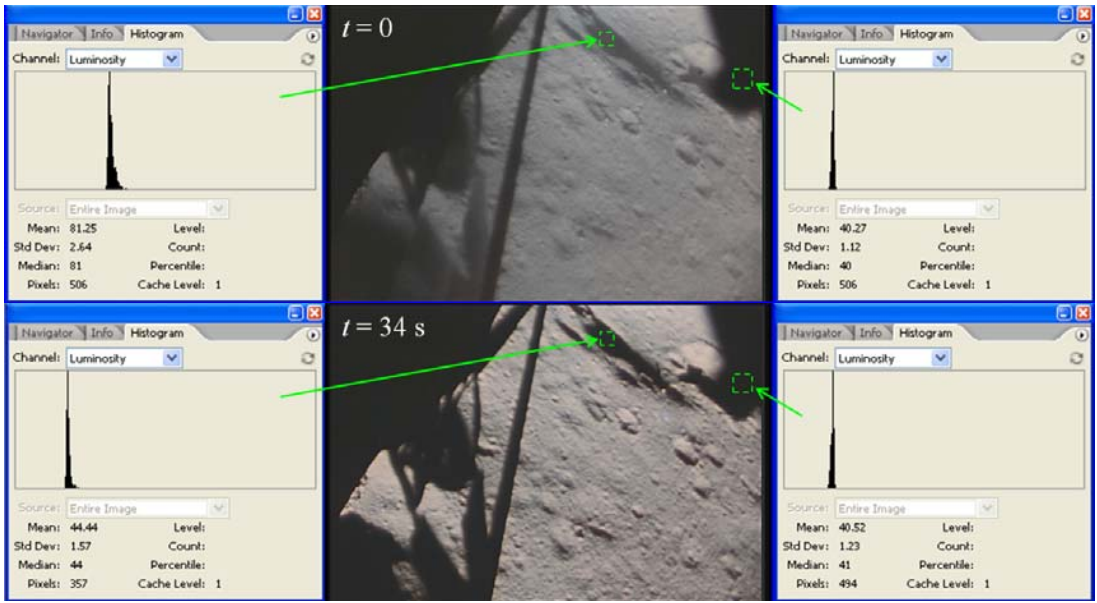

Figure 9: Luminosity measurements of Apollo 14 landing videos following engine cutoff.

The terminal velocity term is due to the drag on the particle by the gas and the pull of lunar gravity. Since the gas density is very small as compared to the particle density $\rho_p$, the terminal velocity $v_T(D)$ depends only on gas viscosity $\mu(T)$, as given by Equation (2), as well as gravity $g$ and particle diameter $D$:

$$v_T(D) = \rho_p g D^2 / 18\mu(T) \qquad (19)$$

To demonstrate the ideas described above, Apollo landing videos can be used. Figure 9 shows a view captured by the Apollo 14 cockpit camera immediately after engine cutoff. The luminosity value $L(t)$ is found in two regions of the image at $t = 0$ and again at $t = 34$ s. The right histograms corresponds to a region imaged inside of the LM, part of the window frame. The histograms on the left correspond to a shadowed region on the lunar surface partly obscured by dust. Figure 9 plots the normalized relative luminosity, $L_N(t) = (L(t) - L_\infty)/L_0$, over several frames. In this case, $L_0 = L(0)$ and $L_\infty = L(34)$. For comparison, a dust particle falling freely under lunar gravity from a height $y_0 = 10$ m is shown, where $y(t) = y_0 - g\,t^2/2$. Also plotted in Figure 9 is Equation (18) for several values of particle diameter with a temperature $T = 250$ K, $A = 1$ m², $\Delta L = 0.2$ m, $\kappa = 10$ $\mu$m², $\rho_p = 3100$ kg/m³, and $M = 14$ metric tons.

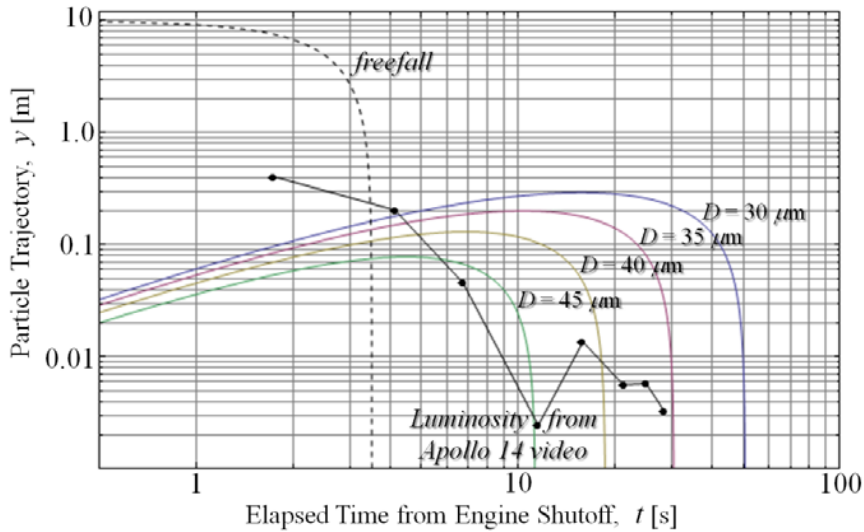

**Figure 10: Freefall and Darcy law particle trajectories compared to luminosity.**

The first conclusion that can immediately be reached by inspection of Figure 10 is that the time required for dust clearing to occur is longer by an order of magnitude then would be explained by freely falling dust under lunar gravity. The same is true for dust under the influence of a static electric field since the electric field term would more likely decrease the clearing time as opposed to an unlikely balance of electric and gravity forces leading to a longer clearing time.

Dust propelled vertically by the Darcy effect could circumstantially explain the dust clearing since there are a range of conditions where Equation (18) matches the clearing time corresponding to the luminosity values measured in the Apollo 14 video. Ignoring the gross assumptions that led to Equation (18), the Darcy Law simulations do not show the dust travelling high from the surface, which for Figure 10 is only in the 10 – 20 cm range for particles in the 20-50 μm range. Smaller particles travel higher. This may however be consistent with the Apollo videos since the main landing dust plume (before engine cutoff) is believed to be contained within a three degree sheath radially centered about the nozzle, which for a distance of 5 m form the nozzle corresponds to dust only up to 25 cm above the surface. The dust viewed by the videos before and after engine cutoff, is primarily seen as a haze over the surface.

## DISCUSSION

X-ray Photoelectron Spectroscopy (XPS) measurements of Surveyor III samples in Table 1 were performed, which is work in progress. Other work in progress includes SEM/EDS analysis of the coupons, as well as particle size distribution analysis of particles on the coupons. Most of the larger particles are no longer on the coupons but are now stuck to the sides of the plastic bags that the Surveyor camera housing samples are stored and transported in. According to some researchers (Wang, 2008) "micrometeoroid impacts and/or disturbances by human activities can also contribute to the liftoff of dust grains from the lunar surface, electrostatic levitation is generally accepted as the primary mechanism.". However, according to Wang's simulations, electrostatic levitation of dust is limited to sub-micron size particles. Based on the XPS and SEM analysis of the sample coupons, the dust coating is not limited to sub-micron size particles so that if electrostatic levitation is an obvious mechanism, it is only one of several other mechanisms.

In previous work (Immer, 2011a) showed that the direct particle ejecta from the LM landing removed more dust from the Surveyor's surface than it deposited. Larger regolith particles were is some cases embedded in the Surveyor, but the smaller particles were wiped clean. This effect is confined to the angular direction of the LM landing site, northeast of the Surveyor III landing site.

Saltation velocity threshold formulas (Sagan, 1990) were implemented using Mathematica and Fortran models and compared to the gas saltation velocity computed from the Fluent CFD (Xiaoyi Li – NASA GRC). Sagan (1990) predicts interparticle cohesion forces on the order of µN for particles in the range of 10 to 100 µm while Walton (2008) predicts a cohesion force about 1000 times larger. Even for the small Sagan cohesion force, particles are not predicted to make it to the Surveyor III site at $R = 109$ m. The actual nozzle height of the LM flyby nearest approach was 67 m, but CFD simulations up to only $h = 45$ were successful. Based on a rough extrapolation estimate, no particles will be lifted in the $h = 67$ m case. However, it is believed that secondary collisions of larger particles with the surface soil may likely lead to smaller particles being ejected and impacting the Surveyor, possibly coating it with a layer of fine dust. It is also important to realize that these predictions are at best useful within a factor of two, so that some particle spray flux on the Surveyor from the flyby may be more likely than the numbers predicted by the simulations. This mechanism is only relevant in the direction of the flyby (north).

Computational work was also done to model the soil pressurization during a lunar landing and the out-gasing from the soil after engine shut down. The time taken to significantly out-gas the soil was predicted for various conditions and range of parameter values. This was compared to the observed dust settling time (~30 seconds) obtained from the Apollo landing videos. These simulations show that the time constant associated with the out-gasing is consistent with the optical opacity decay seen in the Apollo videos. Darcy's law has been shown to partially explain the Apollo landing video dust clearing phenomenon, and therefore partially explain the Surveyor dust coating. This mechanism is an omin-directional effect.

The final suspect is the dust kicked up by the rough landing of the Surveyor. If dust reached the camera housing in this way, electrostatic forces or van der Waals forces would have made it stick as explained by Walton (2008). However, there can be no final verdict on the dust kicking theory without a model that predicts the height that dust will reach and the size distribution of particles levitated in this way.